\documentclass[12pt,preprint]{aastex}

\newcommand{\Msun}{M_\odot}  

\shorttitle{Cluster Formation}
\shortauthors{Huff \& Stahler}

\begin{document}

\title{Cluster Formation in Contracting Molecular Clouds}

\author{E. M. Huff and Steven W. Stahler}
\affil{Astronomy Department, University of California,
    Berkeley, CA 94720}

\email{ehuff@astro.berkeley.edu \,\,  Sstahler@astro.berkeley.edu}

\begin{abstract}
We explore, through a simplified, semi-analytic model, the formation of
dense clusters containing massive stars. The parent cloud spawning the cluster
is represented as an isothermal sphere. This sphere is in near force balance
between self-gravity and turbulent pressure. Self-gravity, mediated by
turbulent dissipation, drives slow contraction of the cloud, eventually leading
to a sharp central spike in density and the onset of dynamical instability. We
suggest that, in a real cloud, this transition marks the late and rapid 
production of massive stars.

We also offer an empirical prescription, akin to the Schmidt law, for 
low-mass star formation in our contracting cloud. Applying this prescription to
the Orion Nebula Cluster, we are able to reproduce the accelerating star 
formation previously inferred from the distribution of member stars in the HR 
diagram. The cloud turns about 10~percent of its mass into low-mass stars before
becoming dynamically unstable. Over a cloud free-fall time, this figure 
drops to 1~percent, consistent with the overall star formation efficiency of
molecular clouds in the Galaxy.

\end{abstract}

\keywords{open clusters and associations: individual (Orion Nebula Cluster) ---
stars: formation --- stars: pre-main-sequence --- ISM: clouds}

\section{Introduction}

There is growing evidence that the formation of stellar groups is a relatively
slow process. More specifically, a star cluster appears within its parent 
molecular cloud over a period long compared to the cloud's free-fall time, 
as gauged by the mean gas density. \citet{tan06} have summarized several lines 
of argument leading to this conclusion. The gas clumps believed to form massive 
clusters look round, indicating that they are in force balance, and not a 
state of collapse. Massive clusters themselves have smooth density profiles, 
again in contrast to a dynamical formation scenario. The observed flux in 
protostellar outflows indicates a slow accretion rate, and therefore a long 
star formation time scale. Finally, placement of young clusters in the HR 
diagram yields age spreads in excess of typical free-fall times (see also 
Palla \& Stahler 2000).  

Many researchers have performed direct numerical simulations of molecular 
clouds; their results also bear on the issue of the star formation time scale.
In a typical simulation, the computational volume is filled with a magnetized,
self-gravitating gas that has a turbulent velocity field. If the turbulence is
only impressed initially, it dies away in a crossing time, and most of the 
gas condenses into unresolvably small structures \citep[e.g.,][]{k98}. Since 
the crossing and free-fall times are similar in a molecular cloud, some authors
have concluded that all clouds produce stars rapidly, while in a state of 
collapse \citep{ha01}. Others have used empirical arguments to make the same 
point \citep{el00}. This view is at odds with the observations concerning
cluster-forming clouds cited above. Moreover, the simulations show that, if 
turbulence is driven throughout the calculation, the rate of star formation 
can be reduced to a more modest level \citep{mk04}.\footnote{The actual rate of 
condensation depends on the magnitude of the {\it sonic length}, i.e., the 
size scale of turbulent eddies whose velocity matches the local sound speed 
\citep{v03,km05}.} It is plausible that the turbulence is indeed driven by 
the cloud's self-gravity, a point we shall amplify later.  

The emerging picture, then, is that molecular clouds both evolve and create
internal clusters in a quasi-static fashion. That is, the structure as a
whole is nearly in force balance, until it is eventually destroyed by the 
ionizing radiation and winds from the very stars it spawns. The inferred masses
of all clouds larger than dense cores greatly exceed the corresponding Jeans 
value, evaluated using the gas kinetic temperature. Thus, self-gravity must be 
opposed by some force beyond the relatively weak thermal pressure gradient. The 
extra support is generally attributed to MHD waves generated by internal, 
turbulent motion \citep{es04}. This motion, which is modeled in the numerical 
simulations just described, imprints itself on molecular line transitions, 
giving them their observed, superthermal width \citep{am75,f92}.

In this paper, we follow the quasi-static contraction of a spherical, 
cluster-forming cloud. Contraction is facilitated by the turbulent dissipation
of energy. This investigation continues and extends an earlier one that was 
part of our study of the Orion Nebula Cluster (ONC) \citep[Paper I]{hs06}. 
Here we track in more detail the changing structure of a generic cloud, taken 
to be in near balance between self-gravity and turbulent pressure. We find that
contraction eventually causes the density profile to develop a sharp, central 
spike. Such a region of growing density is a plausible environment for the 
birth of massive stars \citep[see][]{sp00}.

We also track, using a simple, empirical prescription, the formation of low-mass
stars in our contracting cloud. For reasonable parameter values, stellar births
occur throughout the cloud over a period of order $10^7$~yr. The global rate of
star formation rises with time monotonically, i.e., the formation 
accelerates. Extended, accelerating production of stars is also found 
empirically when one analyzes clusters in the HR diagram \citep{ps00}. Indeed,
it is not difficult to match specifically the global acceleration documented 
in the ONC. Here, the star formation rate depends on cloud density in the same
manner as the classic Schmidt law.

In Section~2 below, we present the basic physical assumptions underlying our 
model. We also give a convenient, nondimensional scheme. In Section~3, we 
introduce our treatment of turbulent dissipation, and calculate the interior
evolution of the cloud as it contracts toward the high-density state. Section~4
offers our prescription for low-mass star formation, and compares the resulting 
birthrate with the ONC data. Finally, Section~5 discusses the broader 
implications of our findings, as well as their utility for future work.     

\section{Formulation of the Problem}
\subsection{Physical Assumptions} \label{assumptions} 

We focus on molecular cloud clumps that are destined to produce the 
highest-density clusters, i.e., those containing massive stars near their
centers. \citet{sh03} used CS observations to study a sample of 63 clumps
already containing massive stars, as evidenced by water maser emission. These
clouds are nearly round, with median projected axis ratios of 1.2. It is
thus a reasonable approximation, and certainly a computationally advantageous
one, to take our model cloud to be spherically symmetric.

The clumps observed by \citet{sh03} have a median radius of 0.32~pc and mass
of $920~\Msun$. A cloud of this size and mass has internal, turbulent motion
well in excess of the sound speed, where the latter is based on the typical
gas kinetic temperature of 10~K \citep{l81}. This bulk motion excites a 
spectrum of MHD waves, i.e., perturbations to the interstellar magnetic field 
threading the cloud \citep{fp86}. Such waves exert an effective pressure that
can, at least in principle, support the cloud against global collapse 
\citep{p90}. 

\citet{fa93} studied the mechanical forcing due to MHD waves propagating in a
one-dimensional, self-gravitating slab. They considered two cases: a slab with 
an embedded magnetic field oriented parallel to the slab plane, and one
with an internal field in the normal direction. In the first case, Fatuzzo
\& Adams showed that magnetosonic waves provide a normal force. In the
second, it is Alfv\'en waves that exert the force, also in the normal direction.
Thus, Fattuzzo \& Adams verified explicitly that the waves counteract gravity,
even in the absence of wave damping.

\cite{mz95} extended this result. Using the pioneering analysis of \citet{d70},
they showed that Alfv\'en waves generated by a turbulent wave field exert an
isotropic pressure, regardless of the background geometry. McKee \& Zweibel 
derived a simple dependence of the wave pressure $P$ on the local density:  
\begin{equation}
P\,\propto\,\rho^{1/2}\,\,.
\end{equation}
If we ignore the relatively small thermal pressure, then the cloud can be
described as an \hbox{$n = -2$} polytrope. 

Are the structures of real clouds consistent with this polytropic wave pressure?
One, indirect, argument indicates they are not. McKee \& Zweibel also 
demonstrated that $P$ is proportional to $\rho$ times the square of the
(randomly oriented) velocity fluctuation $\delta v$. It follows that
\begin{equation}
\delta v \,\propto\, \rho^{-1/4}
\end{equation}
in this model. Now the density in an \hbox{$n\,=\,-2$} polytrope tends to
approach a power law outside the central plateau, such that $\rho$ is
proportional to $r^{-4/3}$. From equation (2), it follows that $\delta v$ is
proportional to $r^{1/3}$. 

Consider the nearly spherical cloud, now gone, that produced the ONC. This cloud
was recently driven off by the Trapezium stars, which themselves have ages of
about $10^5$~yr \citep{ps01}. The disruption itself occurred well within the 
cluster crossing time of about $10^6$~yr. Hence, the present-day velocity 
dispersion of the {\it stars} should reflect the prior $\delta v$ of the 
{\it gas}. But the dispersion of ONC proper motions has negligible variation 
from the center to the outskirts of the cluster \citep{jw88}. These 
measurements span at least a decade in radius, over which $\delta v$ should 
vary by a factor of 2.2, according to the polytropic relation.

Our conclusion, based on this admittedly limited evidence, is that a more
realistic model of the internal turbulence has a spatially constant velocity 
dispersion.\footnote{Inside giant cloud complexes, the observed velocity 
dispersion increases with the size of the substructure \citep{om02}.
Again, we are focusing on a single clump, where such considerations do not
apply.} If we further appropriate the relationship between $P$ and
$\delta v$ derived by McKee \& Zweibel, we are then positing an isothermal
equation of state:
\begin{equation}
P\,=\,\rho\,a_T^2
\end{equation}

Here $a_T$, the effective isothermal sound speed, is taken to be a fixed
constant at a given instant of time. This same quantity varies temporally; 
indeed, this latter variation essentially drives the cloud's evolution. We
emphasize that $a_T$ does not, as in ordinary gas dynamics, give the 
magnitude of random, microsopic velocities. Instead, this quantity
represents, however crudely, the bulk motion of turbulent eddies; these 
eddies create the pressure $P$ via MHD waves. 

Since we are modeling the cloud as an isothermal sphere, we face the familiar
difficulty that its mass is infinite unless the configuration is bounded
externally. We therefore picture the cloud as being surrounded by a low-density,
high-temperature medium with an associated pressure $P_\circ$. This latter
quantity is also the pressure at the boundary of our spherical cloud. The cloud
density at the boundary, $\rho_\circ$, is found from equation (3), given
knowledge of $a_T^2$.

\subsection{Nondimensional Scheme} \label{nondim}

The mathematical description of a self-gravitating, isothermal cloud in
hydrostatic balance is well known \citep[see][Chap.~9]{sp04}. All
structural properties follow from the isothermal Lane-Emden equation:
\begin{equation}
{1\over{\xi^2}}\,{{d{\phantom \xi}}\over{d\xi}}\! 
\left(\xi^2\,{{d\psi}\over{d\xi}}\right)\,=\,{\rm exp}\,(-\psi)\,\,,
\end{equation} 
with boundary conditions \hbox{$\psi(0)\,=\,\psi^\prime(0)\,=\,0$}. Here,
$\psi$ is the dimensionless form of the gravitational potential $\phi_g$: 
\begin{equation}
\psi \,\equiv\, \phi_g/ a_T^2 \,\,.
\end{equation}
The nondimensional radius $\xi$ is obtained from the dimensional $r$
using $G$, $a_T^2$, and the central density $\rho_c$:
\begin{equation}
\xi\,\equiv\,\left({{4\,\pi\,G\,\rho_c}\over{a_T^2}}\right)^{1/2} r\,\,.
\end{equation} 

Equation (4) was derived using both Poisson's equation and the condition
of hydrostatic equilibrium. The latter may be recast as a relation between the
density at any radius, $\rho$, its central value, $\rho_c$, and the potential:
\begin{equation}
\rho \, =\, \rho_c\,{\rm exp}\left(-\psi\right)\,\,.
\end{equation}
The full, dimensional mass $M_\circ$ follows by integration of $\rho$ over mass
shells, Using equation (4) to evaluate the integral, one finds
\begin{equation}
M_\circ \,=\, {a_T^3\over{\sqrt{4\,\pi\,\rho_c\,G^3}}}
\left(\xi^2\,{d\psi\over{d\xi}}\right)_\circ \,\,.
\end{equation}
Here, the subscript denotes the cloud boundary. Similarly, we shall
use $R_\circ$ for the radius of that point, where the internal cloud
pressure has fallen to $P_\circ$.

In this standard formulation, all nondimensional variables are
defined through the basic quantities $a_T^2$, $\rho_c$, and $G$.
Although the standard variables will remain useful, the scheme itself
is not well suited to describing cloud evolution at fixed mass. For
this purpose, we shall also utilize a second nondimensional scheme,
based on $M_\circ$, $P_\circ$, and $G$. 

Let $\lambda$ be the new nondimensional radius, and $\alpha$ the
nondimensional effective sound speed. These are defined through
\begin{equation}
\lambda \,\equiv\,
{{P_\circ^{1/4}}\over{G^{1/4}\,M_\circ^{1/2}}}\ r \,\,\,,
\end{equation}
and
\begin{equation}
\alpha^2 \,\equiv\,{a_T^2\over{G^{3/4}\,M_\circ^{1/2}
\,P_\circ^{1/4}}} \,\,.
\end{equation} 
We further denote as $\delta$ the nondimensional density:
\begin{equation}
\delta \,\equiv\,{{G^{3/4}\,M_\circ^{1/2}}
\over P_\circ^{3/4}}\ \rho \,\,\,.
\end{equation}
Since we will be discussing temporal evolution, we define a 
nondimensional time through
\begin{equation}
\tau \,\equiv\,{{G^{1/8}\,P_\circ^{3/8}}
\over M_\circ^{1/4}}\ t \,\,\,.
\end{equation}

It will be useful to relate new nondimensional quantities to
old ones. Thus, equation (6) tells us $\lambda$ as a function
of $\xi$:
\begin{equation}
\lambda \,=\, \sqrt{\alpha^2\over{4\,\pi\,\delta_c}}\ \xi \,\,.
\end{equation}
For the central density appearing here, $\delta_c$, we use
equation (7), evaluated at the cloud boundary:
\begin{equation}
\delta_c \,=\, {
{{\rm exp}\left(\psi_\circ\right)}\over\alpha^2} \,\,.
\end{equation}
Finally, $\alpha$ itself may be written in terms of standard
variables by using equation (8):
\begin{equation}
\alpha^4 \,=\, \sqrt{4\,\pi}
\left(\xi^2\,{{d\psi}\over{d\xi}}\right)_\circ^{-1}
{\rm exp}\left(\psi_\circ/2\right) \,\,.
\end{equation}

\section{Cloud Evolution}

\subsection{Internal Structure \label{internal}}

We now consider a sequence of isothermal spheres of fixed mass, all embedded in
the same external pressure. We may describe each structure using the new,
nondimensional variables. The sequence is characterized by a single parameter,
the center-to-edge density contrast; we shall denote this ratio as $\beta$. 
From equation (7), $\beta$ can also be written as
\begin{equation}
\beta \,=\, {\rm exp}\left(\psi_\circ\right)
\end{equation}
Since $\psi (\xi)$ is a known function, there is a one-to-one correspondence
between our fundamental parameter $\beta$ and $\xi_\circ$, the old,
nondimensional radius. The potential $\psi$ increases monotonically with $\xi$,
so $\beta$ likewise increases with $\xi_\circ$. The lowest value of $\beta$ is
unity, corresponding to \hbox{$\xi_\circ \,=\, 0$}.

The internal velocity dispersion $\alpha$ varies along our sequence. We may
track this change through equation (15). Thus, for each selected $\beta$, we
first find $\psi_\circ$ from equation (16). From knowledge of the function
$\psi (\xi)$, we find the corresponding $\xi_\circ$, as well as
$\left(d\psi/d\xi\right)_\circ$. Equation (15) then yields $\alpha$.

It is equally straightforward to obtain the internal density profile,
$\delta (\lambda)$, of any model. Knowing $\psi_\circ$ and $\alpha$, 
equation (14) gives the central density, $\delta_c$. Proceeding outward,
equation (13) gives the value of $\xi$ corresponding to each $\lambda$. Again
using $\psi (\xi)$, equation (7) yields the density ratio,
\hbox{$\delta/\delta_c \,=\, {\rm exp}(\psi)$}. When we get to the edge,
\hbox{$\lambda \,=\, \lambda_\circ$}, we find that
\hbox{$\delta_\circ/\delta_c \,\equiv\, \beta^{-1} \,=\,
{\rm exp}(-\psi_\circ)$}, in agreement with equation (16). 

Figure~1 displays graphically the change of the cloud's structure as a function
of $\beta$. Here we have plotted the radius, $\lambda (\beta)$, of selected mass
shells. As expected, a shell in the deep interior monotonically shrinks. Other
shells, however, turn around. With rising $\beta$, an increasing fraction of
the cloud mass starts to expand. Such expansion costs energy. Thus,
configurations of very high $\beta$ are not physically accessible, as we shall
see.  

\subsection{Enthalpy and Dynamical Stability}

The lower dashed, horizontal line in Figure~1 corresponds to a $\beta$-value of
14.1. This is the Bonnor-Ebert density contrast. In the standard analysis, 
clouds of higher contrast are dynamically unstable \citep{e55,b56}. We recall,
however, that this instability arises from perturbations of a cloud {\it held 
at fixed temperature}. In contrast, our sequence has varying effective sound 
speed. The Bonner-Ebert contrast no longer marks a stability transition. 
Nevertheless, this value, which we denote as $\beta_{\rm min}$, is still of 
interest. It signifies, at least in an approximate way, the point where 
self-gravity starts to overwhelm external pressure as the main compressive 
force. Our description of cloud evolution will henceforth focus on such 
gravity-dominated configurations, i.e., those for which 
\hbox{$\beta\,>\,\beta_{\rm min}$}.

To analyze stability in the present sequence, we first need to invoke
thermodynamics. We showed in Paper~I that energy dissipation in an isothermal
cloud results in a decrease of the total enthalpy. Returning to dimensional
notation, equation (A10) stated
\begin{equation}
{{dH}\over{dt}} \,=\, -L \,\,,
\end{equation}
where $L$ is the luminosity. The enthalpy $H$ is the generalization, to a
self-gravitating gas, of the classic definition:
\begin{equation}
H \,\equiv\, E_{\rm therm} \,+\, E_{\rm grav} \,+\, P_\circ\,V \,\,,
\end{equation}
where $E_{\rm therm}$ and $E_{\rm grav}$ are the thermal and gravitational
potential energies, respectively, and $V$ is the cloud volume. 

To evaluate $E_{\rm therm}$, we employ the general expression for a
nonrelativistic gas, $(3/2)\int\!P\,dV$. Using equation (3) for $P$, this
integral becomes $(3/2) M_\circ\,a_T^2$. Instead of evaluating $E_{\rm grav}$
directly, we invoke the virial theorem, in the form
\begin{equation}
E_{\rm grav} \,=\, -2\,E_{\rm therm} \,+\,3\,P_\circ\,V \,\,.
\end{equation}
After expressing the cloud volume in terms of the radius, the enthalpy is 
\begin{equation}
H \,=\, -{3\over 2}\,M_\circ\,a_T^2 \,+\, 
        {{16\,\pi}\over 3}\,P_\circ\,R_\circ^3 \,\,.
\end{equation}
If we define a nondimensional enthalpy $h$ through
\begin{equation}
h \,\equiv\, {H \over{G^{3/4}\,M_\circ^{3/2}\,P_\circ^{1/4}}} \,\,,
\end{equation}\
then equation (20) becomes
\begin{equation}
h \,=\, -{3\over 2}\,\alpha^2 \,+\, 
         {{16\,\pi}\over 3}\,\lambda_\circ^3 \,\,.
\end{equation}

Figure~2 shows $h$ along our sequence of clouds. Again, we restrict ourselves
to gravity-dominated configurations, for which 
\hbox{$\beta \,>\, \beta_{\rm min}$}. We also recall that $\beta$ increases
monotonically along the sequence. Plotted here against $\alpha$, the enthalpy
dips to a minimum, then spirals inward toward a point. The latter corresponds
to the singular isothermal sphere. For this special configuration, it may be
shown that \hbox{$\alpha^2\,=\,(\pi/2)^{1/4}$} and 
\hbox{$\lambda_\circ \,=\, ({1/{8\pi}})^{1/4}$}. Thus, the limiting 
value of $h$ is $-0.187$.

However, this limiting value is never reached in the course of evolution. As
long as the cloud releases energy into space, so that \hbox{$L\,>\,0$}, 
equation~(17) tells us that the enthalpy declines. Thus, the last accessible
configuration coincides with the minimum-enthalpy point in Figure~2. 
Numerically, we find that \hbox{$h\,=\,-0.50$} for this cloud. The corresponding
density contrast $\beta$ is 370.

Consider now two configurations with identical values of $h$, very slightly
above the minimum. These clouds, like all those in the sequence, have the same
mass. We may view them as extremal states attained by the minimum-enthalpy
cloud in the course of a normal mode of oscillation. Here, we are assuming that
the cloud radiates negligible energy during an oscillation period, so that $h$
remains constant. The two endstates are in precise force balance; intermediate
states depart only slightly from this condition. In the small-amplitude limit,
the oscillation has zero frequency, and the unperturbed, minimum-enthalpy, 
state represents a stability transition.

In summary, an isothermal cloud becomes dynamically unstable at a density 
contrast $\beta$ of 370, provided the global enthalpy is held fixed 
during any oscillatory perturbation. This important fact was first discovered by
\citet{c03} in the course of a general analysis of isothermal configurations.
\footnote{Chavanis finds a slightly higher critical $\beta$ of 390. His
minimum enthalpy value, in our units, is \hbox{$h\,=\,-0.493$}.} Note again the 
marked contrast with the traditional, Bonnor-Ebert result. The much lower 
critical density contrast in that case (\hbox{$\beta \,=\, 14.1$}) arises 
because the cloud releases $-$ and draws in $-$ as much energy as necessary to 
remain isothermal, even during a single oscillation period. This assumption 
would be inconsistent with our picture that the cloud is quasi-statically 
contracting due to relatively slow, turbulent dissipation.

The minimum-enthalpy state thus marks the natural endpoint of the cloud's
evolution. We denote as $\beta_{\rm max}$ the corresponding density contrast,
and display this limit as the upper dashed, horizontal line in Figure~1. Clouds
with higher density contrast, including the singular isothermal sphere, are
inaccessible.

\subsection{Turbulent Dissipation}  \label{turb} 

Although we have drawn a number of conclusions regarding the changing structure
of our model cloud, we have yet to discuss its temporal evolution. The
quasi-static contraction envisioned here is facilitated by the release of 
energy. This emission must arise at the shock interface between colliding,
turbulent eddies. Typical fluid speeds are the virial value, i.e., less than
10~km~s$^{-1}$ for the clouds of interest. Hence, the shocks radiate through
far-infrared and submillimeter photons from low-lying transitions of molecules.
The cloud is optically thin to such photons. The luminosity $L$ in 
equation~(17) is thus generated from the full interior.

Consider, then, a representative volume of the cloud. The numerical simulations
mentioned previously have modeled the dynamics of a magnetized gas subject to
an impressed, turbulent velocity field. Even if the fluid disturbances begin as
incompressible Alfv\'en waves, efficient mode conversion produces compressible
MHD waves that steepen and shock \citep[see, e.g.,][]{g78}. Signficant energy is
dissipated during the characteristic crossing time of the largest eddies. For
example, \citet{m99} found that 
\begin{equation}
{\dot\epsilon} \,=\, -\eta\,{V^3_{\rm turb}\over\lambda} \,\,.
\end{equation}
Here $\dot\epsilon$ is the energy loss rate per unit mass of gas, $V_{\rm turb}$
the average (rms) eddy speed, and $\lambda$ the dominate wavelength of the
impressed turbulence. The empirical constant $\eta$ was measured by Mac~Low to
be about 0.4.

In the simulations, turbulence is impressed on an arbitrary scale. Indeed, the
question of what drives the turbulence remains controversial \citep{m04}. Here,
we recall the key fact that the mean, interior velocities match the virial value
over a large range of cloud sizes and masses \citep{l81}. It is likely, 
therefore, that self-gravity constitutes the basic driving mechanism, although
a quantitative model is still lacking. If this basic idea is correct, then the
``dominant wavelength" in equation~(23) should be comparable to the cloud size.
We therefore adopt, as our expression for the cloud luminosity $L$, a 
mass-integrated version of this relation:
\begin{equation}
L \,=\, \eta\,{{M_\circ\,a_T^3}\over {4\,R_\circ}} \,\,,
\end{equation}
where the prefactor $\eta$ does not necessarily have the value found by
Mac~Low. The factor of 4 in the denominator reflects the fact that the 
largest mode corresponds to overall expansion or contraction of the cloud 
\citep{ma02}.

\subsection{Quasi-Static Contraction}

To follow the cloud evolution in time, we use our prescribed luminosity to
alter the global enthalpy. Combining equations (17), (20), and (24), we
recast the result into nondimensional terms:
\begin{equation}
-\eta\,{\alpha^3\over\lambda_\circ}\,=\,
-{6}\,{{d\alpha^2}\over{d\tau}} \,+\,
64\,\pi\lambda_\circ^2\,{{d\lambda_\circ}\over{d\tau}} \,\,.
\end{equation}

The dependent variables $\alpha$ and $\lambda_\circ$ are already known 
implicitly in terms of $\xi_\circ$. (Recall equations (13) $-$ (15).) We may
thus regard equation~(25) as giving the dependence of $\tau$ on this same
quantity. Since $\eta$ is still unknown, we use instead the combination 
$\eta\tau$:    
\begin{equation}
{{d(\eta\tau)}\over{d\xi_\circ}} \,=\, 
{{6\,\lambda_\circ}\over{\alpha^3}}\,{{d\alpha^2}\over{d\xi_\circ}} \,-\,
{{64\,\pi\,\lambda_\circ^3}\over{\alpha^3}}\,
{{d\lambda_\circ}\over{d\xi_\circ}}\,\,.
\end{equation}
We integrate this equation numerically, setting \hbox{$\eta\tau\,=\,0$} at
\hbox{$\xi_\circ\,=\,6.5$}, the value at the Bonnor-Ebert density contrast, 
and ending at \hbox{$\xi_\circ\,=\,25$}, the minimum-enthalpy state. Over this
evolutionary span, $\eta\tau$ increases by 0.96.

Figure~3 shows that neither the effective sound speed nor the cloud radius
vary greatly during this period. The former increases monotonically, with a
fractional change of 10~percent by the end. The radius gently decreases most of
the time. Just before the unstable state is reached, the cloud surface begins to
swell, in agreement with Figure~1.

The temporal change of the central density is much more dramatic. As seen in
Figure~4, $\delta_c$ increases slowly at first, and then accelerates strongly 
at the end. Given the behavior of the cloud radius, this rapid compression
evidently involves a small fraction of the total cloud volume. The left panel
of Figure~5 displays the evolution of the full density profile. From bottom to
top, the associated values of $\eta\tau$ are 0, 0.15, 0.23, and the final
value, \hbox{$\eta\tau \,=\, 0.24$}.

Because the inner portion of the cloud undergoes such rapid compression, one 
may question the basic assumption of quasi-static behavior. Are all mass shells
really moving at subsonic velocity? At the very center, the velocity must fall
to zero at each time. But what of shells just outside the center?

Figure~6 shows the velocity profiles for the final three times depicted in the
left panel of Figure~5. Since the temporal variable in equation~(26) is
$\eta\tau$, we plot the ratio $v/\eta\alpha$. For \hbox{$\eta\tau \,=\,0.60$}
and 0.92, the velocity is subsonic throughout, even if $\eta$ were unity. In
fact, plausible $\eta$-values are less than unity, as indicated earlier; we
shall see in Section~4 below that \hbox{$\eta\,\approx\,0.3$} best matches the
ONC data. In the final profile, corresponding to the minimum-enthalpy state, 
the peak velocities occur in the deep interior, and are mildly supersonic 
(\hbox{$v\,\gtrsim\,\alpha$}). 

\section{Modeling the ONC}

\subsection{Prescription for Star Formation}

At present, we have scant knowledge of how the birthrate of stars scales with 
the properties of the cloud medium spawning these objects. Within the solar
neighborhood, \citet{s59} found the star formation rate to be proportional to
the square of the local density; here, the rate is measure per unit volume.
Schmidt's law thus states that the formation rate per gas mass scales
linearly with the cloud density.\footnote{In Schmidt's original formulation,
the gas in question was HI. We now know, of course, that the relevant clouds
are molecular.}

On the scale of galactic disks, it is established that the formation rate per 
unit disk area rises as $\Sigma^n$, where $\Sigma$ is the total gas surface 
density, and the exponent $n$ is about 1.4 \citep{ke98}. However, it is not
straighforward to relate this important finding to the present study. Each 
areal patch in the galactic observations comprises numerous molecular 
complexes, any one of which is far larger than the clumps of direct interest 
here.

Returning to our model, if the clump indeed undergoes slow contraction, then 
its overall star formation rate must increase with time. It is plausible that
the local rate within each mass shell rises with that shell's density $\rho$. 
Following Schmidt, we posit a power-law dependence: 
\begin{equation}
{\dot m}_\ast \,=\, {\epsilon\over t_1}\,\left({\rho\over\rho_1}\right)^n
\,\, .  
\end{equation}
Here, ${\dot m}_\ast$ is the mass in stars forming per unit time, per unit 
cloud mass. The fiducial density and time, $\rho_1$ and $t_1$, are those from
equations (11) and (12), respectively: 
\begin{mathletters}
\begin{eqnarray}
\rho_1 \, & \equiv & \, {P_\circ^{3/4}\over{G^{3/4}\,M_\circ^{1/2}}} \\
t_1 \, & \equiv & \, {M_\circ^{1/4}\over{G^{1/8}\,P_\circ^{3/8}}} \,\,.
\end{eqnarray}
\end{mathletters}
Finally, the exponent $n$ is to be set by matching to observations. We stress
that the prescription in equation (27) applies to low-mass stars. The formation 
of massive objects is a separate phenomenon. In our model, this occurs only in
the high-density, central region of the final, minimum-enthalpy state.

Equation~(27) contains a nondimensional efficiency factor $\epsilon$. Since 
only a fraction of the cloud mass turns into stars, we expect $\epsilon$ to be
well under unity. If, for simplicity, we assume this parameter to be the same 
in all mass shells, then integration of equation~(27) yields the total mass per
unit time in new stars:
\begin{equation}
{\dot M_\ast} \,=\, {{4\,\pi\,\epsilon}\over{t_1\,\rho_1^n}} 
\int_0^{R_\circ} \! \rho^{n+1}\,r^2\,dr \,\,.
\end{equation} 
We may conveniently recast this fornula in terms of the effective sound speed:
\begin{equation}
{\dot M_\ast} \,=\, \epsilon\,{a_T^3\over G}\ {\cal I} \,\,.
\end{equation}
Here, the nondimensional quantity ${\cal I}$ is expressed using the 
traditional, polytropic variables:
\begin{equation}
{\cal I} \,\equiv\,  {{\alpha^{1-2n}}\over{\sqrt{4\pi}}}
{\rm e}^{(n-1/2)\psi_\circ} 
\int_0^{\xi_\circ}\!{\rm e}^{-(n+1)\psi}\,\xi^2\,d\xi \,\,.
\end{equation}

Our prescription gives a finite star formation rate for clouds of arbitrarily
low density. This is clearly an oversimplification. There is no evidence, for
example, that HI clouds form stars at all. Even within the molecular domain,
it may be that stars form only above some threshold density. In their study of 
the Rosette cloud complex, \citet{w95} found that only clumps which are 
strongly self-gravitating (as assessed by comparison of velocity dispersions, 
masses, and sizes) have internal stars. While a more complete model should
account for this threshold effect, we shall not include it explicitly, but
simply limit our discussion to self-gravitating clumps
(\hbox{$\beta\,>\,\beta_{\rm min}$}) that are capable of forming stars. 

\subsection{Comparison with Observations}

In Paper I, we empirically determined the star formation history of the ONC. The
database of \citet{h97}, together with theoretical pre-main-sequence tracks
\citep{ps99}, allowed us to assign masses and contraction ages.\footnote
{Recently, \citet{j07} has redetermined the ONC distance as 390~pc, rather
than the 470~pc used by \citet{h97}. If correct, this distance shift will
systematically lower stellar luminosities, and therefore increase their ages.}
In the detailed analysis, we restricted our attention to the 244 members with
\hbox{$M_\ast > 0.4\ \Msun$}; the Trapezium stars themselves have
\hbox{$M_\ast \geq 7\ \Msun$}, and are thus already on the main sequence.
This sample is statistically complete, in the sense that the oldest stars do 
not fall below the observational sensitivity limit.

Based on our results from Paper~I, Figure~7 shows ${\dot M}_\ast(t)$, the mass
production per unit time, as a function of stellar age. Here, we have binned 
the data in age intervals of $10^6$~yr. We have also extrapolated from our 
subsample to all stellar masses. We did so by multiplying the accumulated mass
at each epoch by a factor of 1.3. This factor accounts for the missing stars 
with \hbox{$M_\ast\,<\,0.4\ \Msun$}, according to the field-star initial mass 
function of \citet{s98}. 

In order to use equation (30) to describe the ONC, we need the effective sound
speed $a_T$ as a function of time. Our numerical model gives the nondimensional
functional relation $\alpha (\eta\tau)$ (recall Fig. 3). Similarly, the 
quantity ${\cal I}$ contains $\psi_\circ$ and $\xi_\circ$, which we also know
as functions of $\eta\tau$. Converting these relations to dimensional form
requires that we set the cloud mass $M_\circ$ and background pressure $P_\circ$.

We now make the critical assumption that the parent cloud of the ONC was, just 
prior to its dispersal, in the minimum-enthalpy state. Then equations (9) and
(10) may be combined to yield
\begin{equation}
M_\circ \,=\, f_1\, {{R_\circ\,a_T^2}\over G} 
\end{equation}
and
\begin{equation}
P_\circ \,=\, f_2\, {{G\,M_\circ^2}\over R_\circ^4} \,\,,
\end{equation}
where $R_\circ$ and $a_T$ refer to the final cloud state. The results of our
numerical integration give \hbox{$f_1 \,=\,2.0$} and \hbox{$f_2 \,=\,0.028$}. We
take $R_\circ$ to equal the radius of the stellar cluster (2.5~pc; see
Hillenbrand 1997), and identify $a_T$ with the observed {\it stellar} velocity 
dispersion (2.4~km~s$^{-1}$; see Jones \& Walker 1988). We then find that 
\hbox{$M_\circ \,=\,6900\ \Msun$} and 
\hbox{$P_\circ\,=\,1.1\times 10^{-10}\ {\rm dyne~cm}^{-2}$}. The latter is 
about 300 times the canonical value in the diffuse interstellar medium, i.e., 
that bounding HI clouds \citep{wo95}.  

It remains only to adjust $\epsilon$, $\eta$, and the exponent $n$, until the
theoretical star formation rate ${\dot M}_\ast$, as given by equation~(30), 
matches the empirical one. We use a standard implementation of the 
Levenberg-Marquardt fitting algorithm \citep[  \S~15.5]{num88}. The likelihood 
function that we maximize incorporates a uniform error in the star formation 
rate at each epoch of $\sigma\,=\,3\ \Msun\ {\rm Myr}^{-1}$. Here we have 
used \hbox{$\sigma \,\approx\,\sqrt{N}\langle M\rangle$}, where $N$ is the
median number of stars produced per $10^6$~yr, and \hbox{$\langle M\rangle$}
is the average stellar mass at the appropriate epoch. This formula assumes that
the number of stars in each bin is Poisson distributed about the mean predicted
by the model, i.e., we neglect the observational contribution to the error.  

The dashed curve in Figure~7 shows our theoretical rate as a function of 
stellar age, along with the optimal values for the three parameters. As 
predicted, $\epsilon$ is small ($2\times 10^{-4}$), signifying a low efficiency
for stellar production. Specifically, the cloud converts 8~percent of its mass
($550~\Msun$) into low-mass stars. \footnote{This figure is higher if many of 
the embedded, near-infrared sources seen behind the ONC are members of the 
original cluster \citep{ad95,hc00}.} The parameter $\eta$ is also small (0.3),
indicating that the cloud contracts over an interval long compared with the 
free-fall time, $t_{\rm ff}$. The latter is 1.6~Myr for our initial cloud 
state. Finally, the best-fit value of $n$ (1.4) lies close to unity. Thus, the
original star formation law of Schmidt may hold quite generally. The 1-sigma
errors in $\epsilon$, $\eta$, and $n$, are, respectively, $1\times 10^{-4}$,
$0.08$, and $0.09$. These figures would have been larger had we included
observational sources of error.

We note also that the basic physical characteristics of our model clouds are
consistent with the clump properties inferred from observations. Our fiducial
density $\rho_\circ$ is equivalent to a molecular hydrogen number density of
520~cm$^{-3}$. As can be seen in Figure~5, the average interior density is
higher by an order of magnitude. The mean visual extinction of a cloud, using
$M_\circ/\pi\,R_\circ^2$ as the typical column density, is 11~mag. These
figures are in general accord with the findings of \citet{w95} for 
self-gravitating clumps in the Rosette complex. Moreover, $A_v$, measured
inward from the edge, quickly exceeds unity in all our models. Thus, the
hydrogen is indeed in molecular form throughout the bulk of the interior, as
is appropriate for a star-forming cloud.

\section{Discussion}

In this paper we have adopted a conceptually simple model of cloud contraction
and stellar group formation. The cloud is a self-gravitating sphere, supported
against collapse by the motion of turbulent eddies. We assigned an effective
pressure to this motion; our formulation implicitly assumes the eddy speed to
be spatially constant, though varying in time. Cloud evolution is mediated by
the slow leakage of energy, presumed to occur through internal, shock
dissipation.

This model, supplemented by a Schmidt-type prescription for the star formation,
can account not only for the empirically known history of the ONC, but also 
for more general characteristics of stellar birth. Consider, for example, the 
issue of formation efficiency. In the spirit of \citet{km05}, we may define a 
nondimensional star formation rate per free-fall time:
\begin{equation}
\epsilon_{\rm ff} \,\equiv\, 
{t_{\rm ff}\over t_{\rm ev}} \,
{{\Delta M_\ast}\over M_\circ} \,\,.
\end{equation}
Here, $t_{\rm ev}$ is the time over which the cloud produces stars, while
$\Delta M_\ast$ is the total mass in these objects. Using our ONC model, and
setting \hbox{$t_{\rm ev}\,=\,1\times 10^7~{\rm yr}$}, we find
\hbox{$\epsilon_{\rm ff} \,=\, 0.014$}. \citet{ze74} long ago pointed out that
only about 1~percent of the Galaxy's molecular gas can become stars in a cloud
free-fall time, to reproduce the observed, global star formation rate. The
agreement here suggests that giant complexes create stars principally through
their slowly contracting, internal clumps, as we have modeled.

We have not described, in any quantitative way, the physics underlying the
turbulent dissipation. Our best-fit value of $\eta$ for the ONC is similar to
that found in numerical simulations of turbulent clouds \citep{m99}. However,
our physical picture is quite different. All simulations to date, which focus on
an isolated, interior volume, find the turbulent energy dying away. In our
model, the mean turbulent speed increases with time (see Figure~3). Future,
global simlations of self-gravitating clouds supprted by turbulent pressure 
should show this effect.  

Our best-fit value of $n$ agrees, perhaps fortuitously, with that originally
proposed by \citet{s59}. For an $n$-value of unity, the star formation rate per
cloud mass scales with the gas density. Did this proportionality really hold
in the ONC? The righthand panel of Figure~5 suggests that it did, at least
roughly. Here, the solid curve is the density profile of our final,
minimum-enthalpy, cloud model. The dot-dashed curve is the current number
density of ONC stars, as reconstructed from the observed, projected number
density (see Fig.~3 of Paper~I). The similarity of the two curves indeed
suggests that stars trace the mass distribution of the parent cloud. The same
point is evident when comparing the projected stellar density with CO
contours of the remnant gas \citep[see][Fig.~12.27]{sp04}.

Of course, all stars travel some distance from their birth sites. They do not
move ballistically, but are subject to the gravitational potential of the
parent cloud. Because of the star formation law expressed in equation~(27),
stellar births are indeed concentrated toward the cloud center, but there will
inevitably be some outward diffusion. In a future paper, we hope to track this
process through a direct, numerical simulation.
  
The minimum-enthalpy cloud that terminates our dynamical sequence is dynamically
unstable. How do we interpret this instability in a more realistic setting? The
essential fact is that the self-gravity of the gas becomes so strong that it
leads to rapid, internal contraction, perhaps even true collapse of the central
region. It is tempting to link this event with the formation of high-mass 
stars. While the physics of massive star formation is far from clear, the
collapse or coalescence of dense, gaseous structures appears to play a key
role \citep{sp00,mt03}. In the specific case of the ONC, the Trapezium stars
are, of course, centrally located, and appear to be of relatively recent
origin \citep{ps01}.

In comparing our model with data from the ONC, we have accepted at face value
the stellar ages inferred from the placement of each object in the HR diagram.
The age spread within clusters remains a contentious issue. For the ONC,
\citet{p05} have found that four nominally older stars are depleted in lithium,
as would be expected. Such findings are inconsistent with the assertion by
\citet{h01} that the ostensible age spread primarily reflects observational
uncertainties. 

\citet{pk07} accept the higher ages, but hypothesize that all such stars were 
gravitationally captured from somewhat older, neighboring clusters. As the 
authors themselves note, the existence of such neighboring systems is unclear.
The other subassociations within the Orion complex are too young. One 
possibility is that there were a large number of nearby small groups producing
low-mass stars and then dispersing. Pending more direct evidence for such 
groups, we continue to believe that the nominal age spreads in both the ONC 
and other systems are real.

In the present model, we have taken the cloud to be of fixed mass. This
assumption may be acceptable for the ONC progenitor cloud, at least until
the point when the Trapezium stars ionized and drove off the gas. In clouds
producing low-mass T associations, the latter process does not occur. Yet these
clouds are still dispersed, presumably through stellar winds. In our next
paper, we will generalize our model of cloud contraction to include the effect
of continuous mass loss. We will thus achieve a fuller picture of stellar
group formation, a process of importance not only locally, but on galactic
scales.

\acknowledgments

We are grateful to Steve Shore for pointing out the work of P. Chavanis on
the thermodynamics of self-gravitating spheres. This project was supported by
NSF grant AST-0639743.

\clearpage

\begin{figure}
\epsscale{.80}
\plotone{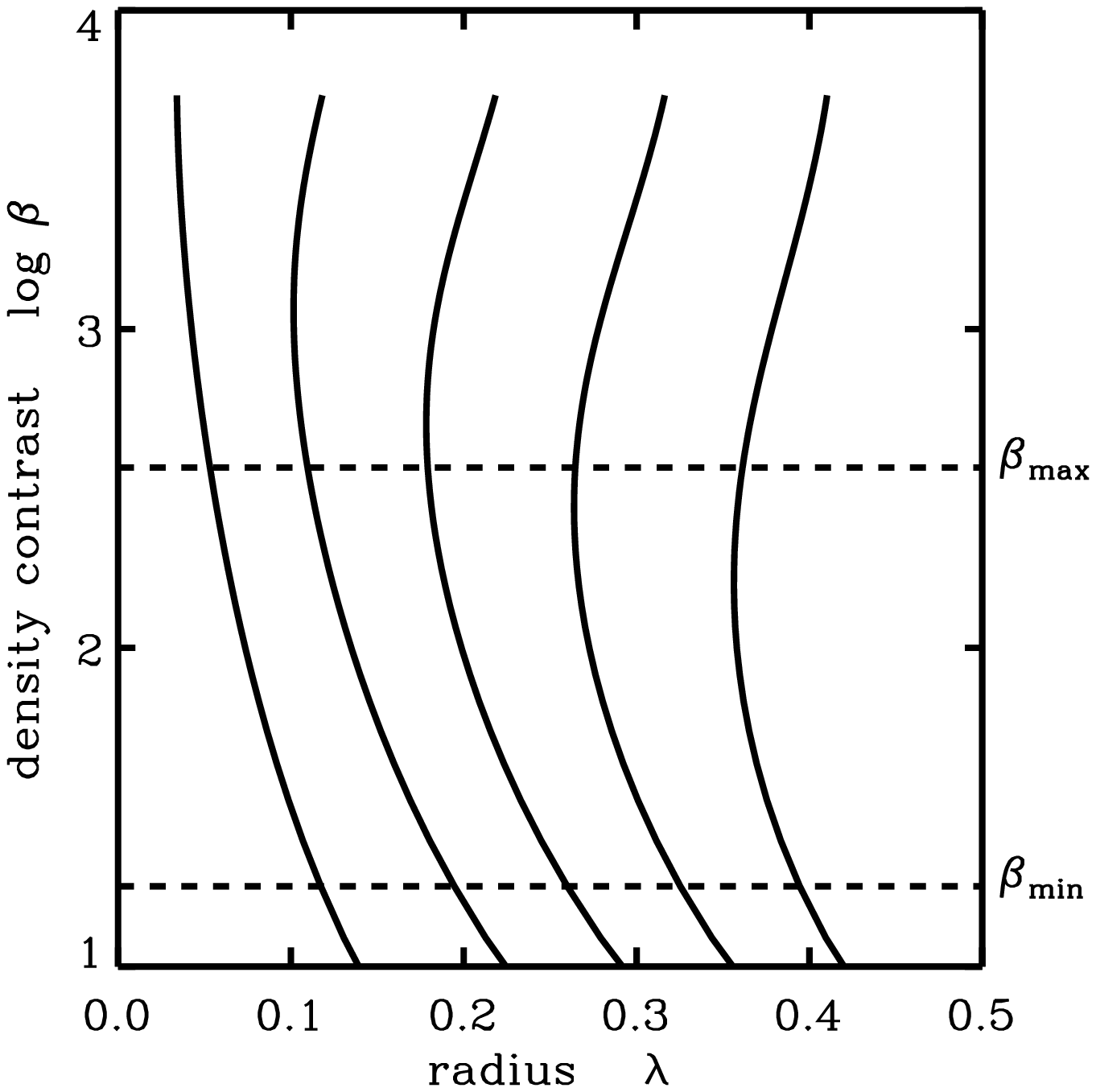}
\caption{Evolution of the cloud's internal structure. Shown are the radii of
selected, Lagrangian mass shells as a function of the density contrast $\beta$.
From left to right, the five shells enclose 0.2, 0.4, 0.6, 0.8, and 1.0 times 
the total cloud mass. Also indicated are the minimum $\beta$-value for 
self-gravitating clouds and the maximum value for dynamical stability.}
\end{figure}
\clearpage

\begin{figure}
\plotone{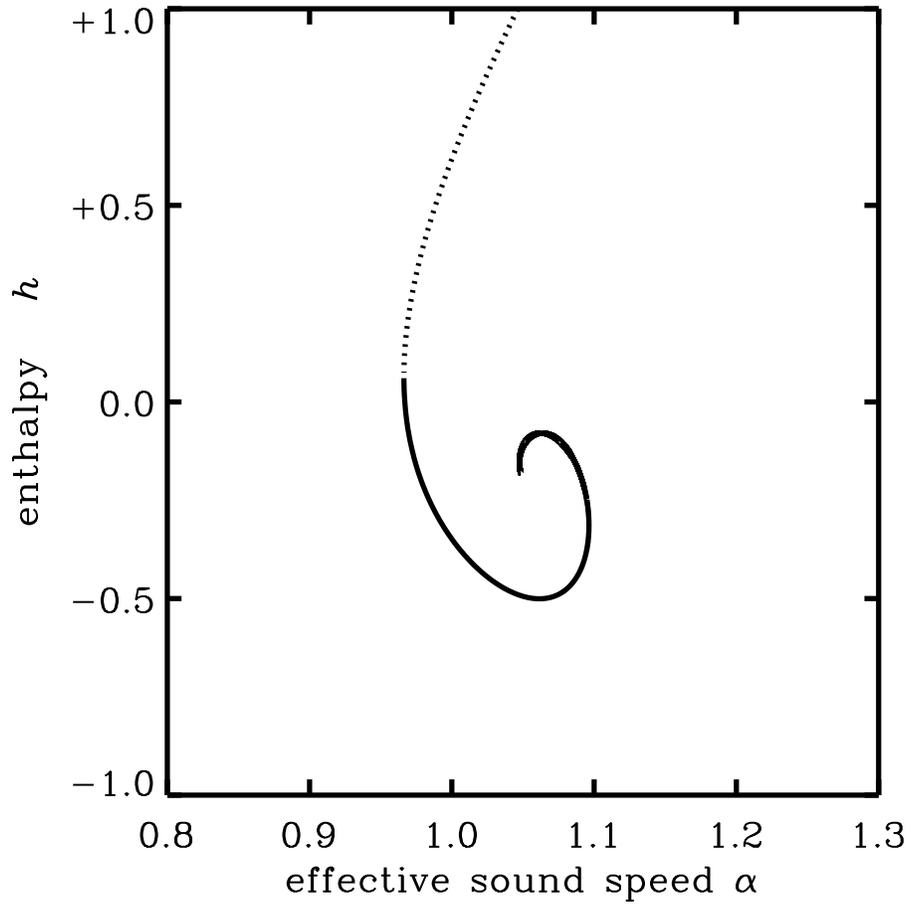}
\caption{The run of specific enthalpy $h$ along the cloud sequence. This
quantity is plotted as a function of the effective sound speed $\alpha$. The
dotted portion of the curve pertains to clouds which have too low a density
contrast to be self-gravitating.} 
\end{figure}
\clearpage

\begin{figure}
\plotone{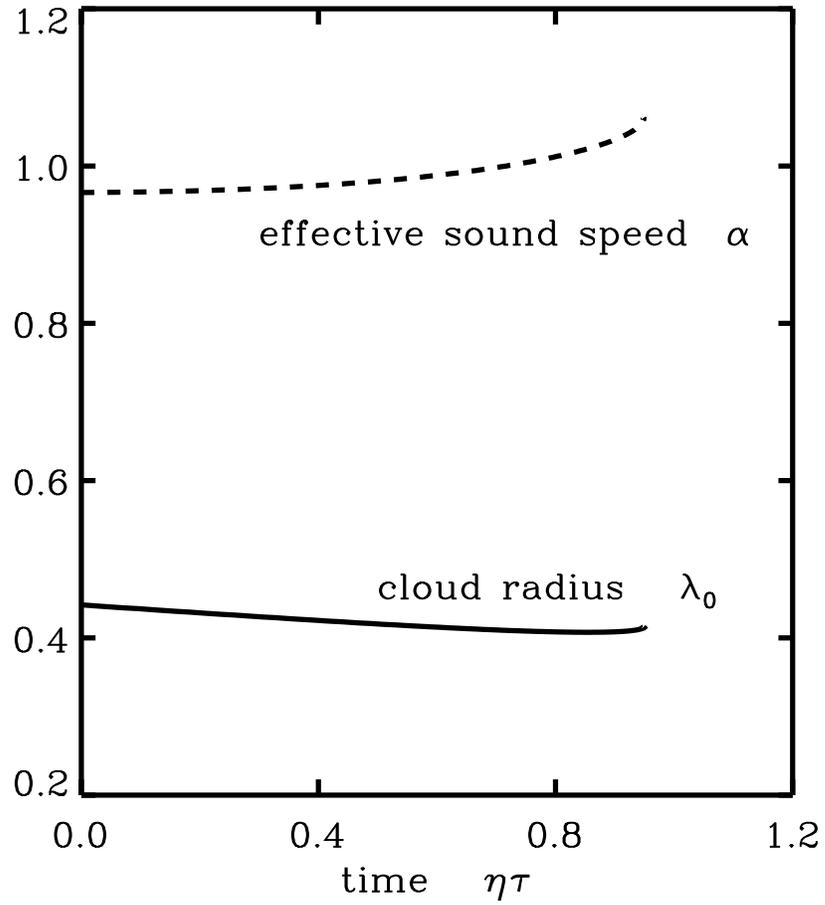}
\caption{Evolution of the cloud radius ({\it solid curve}) and the effective
sound speed ({\it dashed curve}). Notice that the time coordinate $\eta\tau$
starts at the first self-gravitating configuration, i.e., that for which
\hbox{$\beta \,=\, \beta_{\rm min}$}.}
\end{figure}
\clearpage

\begin{figure}
\plotone{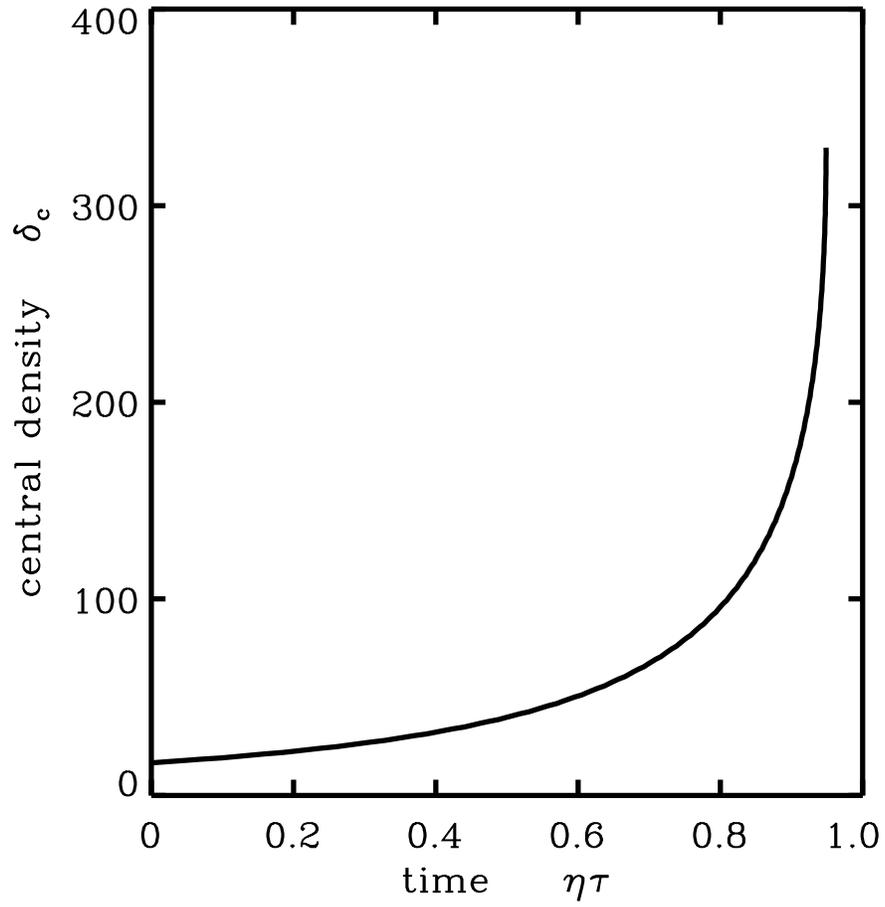}
\caption{Evolution of the nondimensional central density. As in Figure~3, the
time $\eta\tau$ is measured from the first self-gravitating cloud.}
\end{figure}
\clearpage

\begin{figure}
\plotone{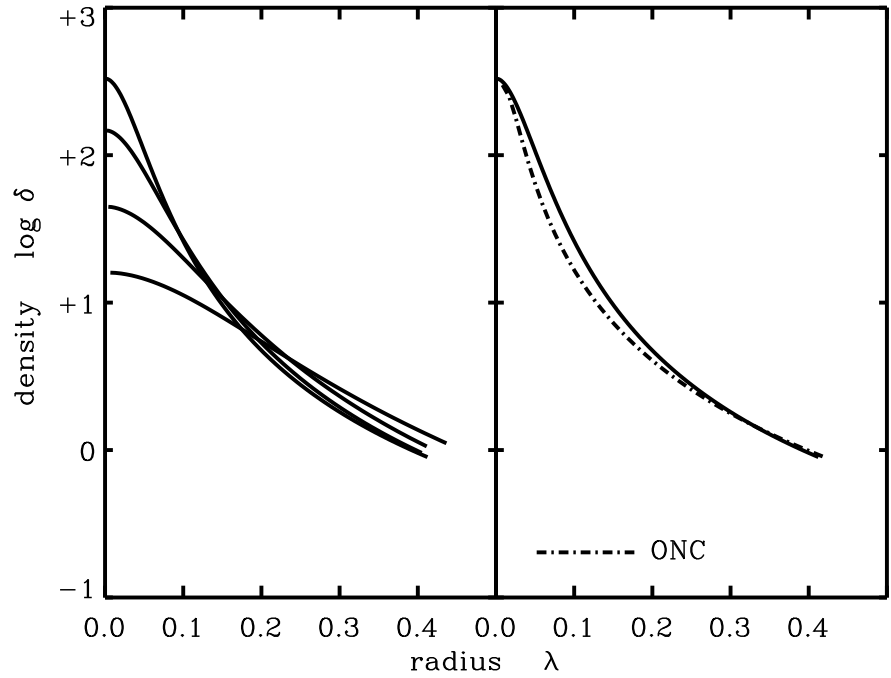}
\caption{{\it Left panel:} Evolution of the density as a function of radius. 
From bottom to top, the corresponding values of $\eta\tau$ are 0, 0.60, 0.92, 
and 0.96. The latter value corresponds to the minimum-enthalpy state. 
{\it Right panel:} The density profile of the minimum-enthalpy state 
({\it solid curve}) compared to the reconstructed stellar number density in the
ONC ({\it dot-dashed curve}).}
\end{figure}
\clearpage

\begin{figure}
\plotone{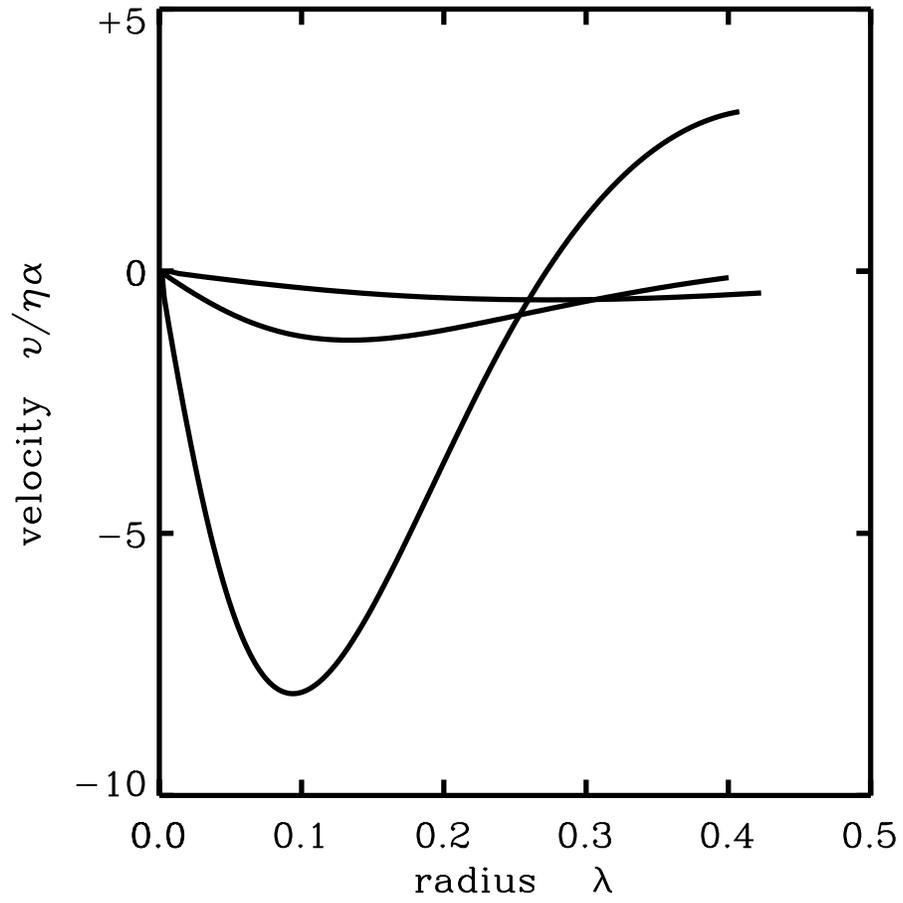}
\caption{Evolution of the velocity profile. In order of
deepening minima, the curves correspond to $\eta\tau$-values of 0.60, 0.92,
and 0.96. Note that the velocity is normalized to $\eta\alpha$.}
\end{figure}
\clearpage

\begin{figure}
\plotone{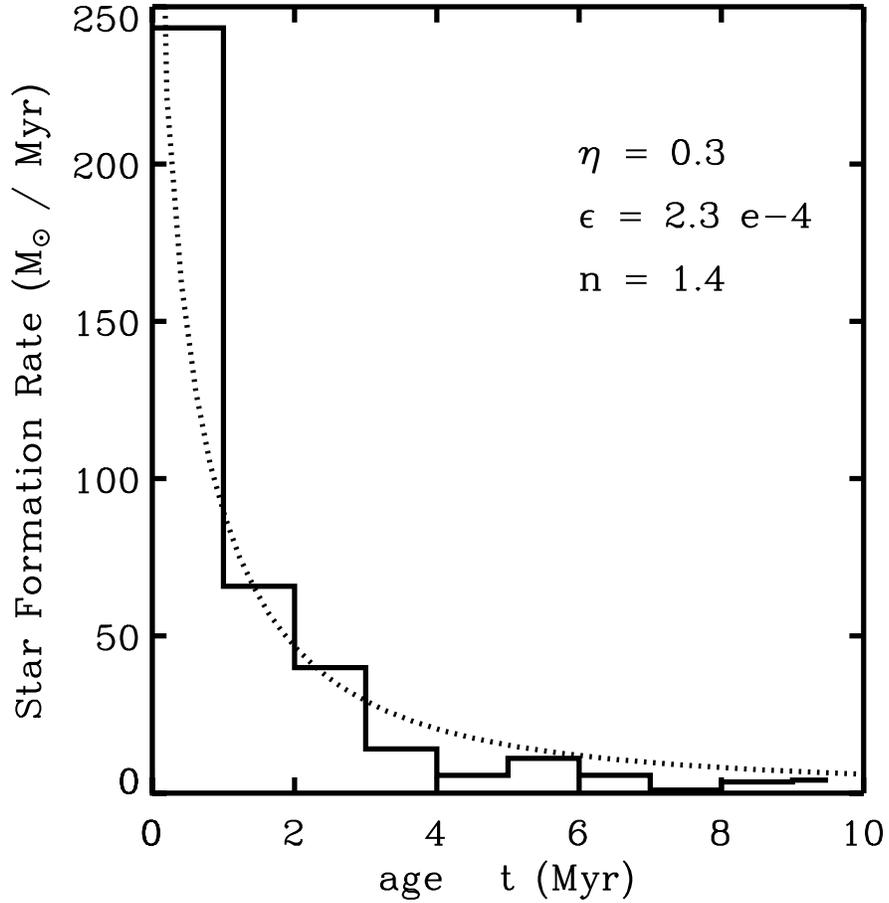}
\caption{Total star formation rate in the ONC as a function of time. The
latter is actually shown as the stellar age. The solid histogram uses the
empirical ages from Paper~I, binned in 1~Myr intervals. The dashed curve is the
theoretical prediction. Also shown are the best-fit values for the model's 
three free parameters: $\eta$, $\epsilon$, and $n$.}
\end{figure} 
\clearpage


\begin{thebibliography}{}
\bibitem[Ali \& Depoy(1995)]{ad95} Ali, B. \& Depoy, D. L. 1995, \aj, 109, 709 
\bibitem[Arons \& Max(1975)]{am75} Arons, J. \& Max, C. E. 1975, \apjl, 196,
    L77
\bibitem[Bonnor(1956)]{b56} Bonnor, W. B. 1956, \mnras, 116, 351
\bibitem[Chavanis(2003)]{c03} Chavanis, P. H. 2003, \aap, 401, 15
\bibitem[Dewar(1970)]{d70} Dewar, R. L. 1970, Phys.Fluids, 13, 2710
\bibitem[Ebert(1955)]{e55} Ebert, R. 1955, \zap, 37, 322
\bibitem[Elmegreen(2000)]{el00} Elmegreen, B. G. 2000, \apj, 530, 277
\bibitem[Elmegreen \& Scalo(2004)]{es04} Elmegreen, B. G. \& Scalo, J. 2004, 
     \araa, 42, 211
\bibitem[Falgarone \& Puget(1986)]{fp86} Falgarone, E. \& Puget, J. L. 1986,
    \aap, 162, 235
\bibitem[Falgarone et al.(1992)]{f92} Falgarone, E., Puget, J.-L., \&
P\'erault, M. 1992, \aap, 257, 730 
\bibitem[Fatuzzo \& Adams(1993)]{fa93} Fatuzzo, M. \& Adams, F. C. 1993,
\apj, 412, 146
\bibitem[Goldstein(1978)]{g78} Goldstein, M. L. 1978, \apj, 219, 700
\bibitem[Hartmann(2001)]{h01} Hartmann, L. 2001, AJ, 121, 1030
\bibitem[Hartmann et al.(2001)]{ha01} Hartmann, L., Ballesteros-Paredes, J., \&
Bergin, E. A. 2001, \apj, 562, 852
\bibitem[Hillenbrand(1997)]{h97} Hillenbrand, L. A. 1997, \aj, 113, 1733
\bibitem[Hillenbrand \& Carpenter(2000)]{hc00} Hillenbrand, L. A. \&
       Carpenter, J. M. 2000, \apj, 540, 236
\bibitem[Huff \& Stahler(2006)]{hs06} Huff, E. M. \& Stahler, S. W. 2006,
       \apj, 644, 355
\bibitem[Jeffries(2007)]{j07} Jeffries, R. D. 2007, astro-ph/0701186
\bibitem[Jones \& Walker(1988)]{jw88} Jones, B. F. \& Walker, M. F.  1988, \aj, 
       95, 1755
\bibitem[Kennicutt(1998)]{ke98} Kennicutt, R. C. 1998, \apj, 498, 541 
\bibitem[Klessen et al.(1998)]{k98} Klessen, R. S., Burkert, A., \&
       Bate, M. R. 1998, \apjl, 501, L205
\bibitem[Krumholz \& McKee(2005)]{km05} Krumholz, M. R. \& McKee, C. F. 2005,
       \apj, 630, 250
\bibitem[Larson(1981)]{l81} Larson, R. B. 1981, \mnras, 194, 809  
\bibitem[Mac~Low(1999)]{m99} Mac~Low, M.-M. 1999, \apj, 524, 169 
\bibitem[Mac~Low(2004)]{m04} Mac~Low, M.-M. 2004, \apss, 289, 323
\bibitem[Mac~Low \& Klessen(2004)]{mk04} Mac~Low, M.-M. \& Klessen, R. S.
       2004, Rev. Mod. Phys., 76, 125
\bibitem[Matzner(2002)]{ma02} Matzner, C. D. 2002, \apj, 566, 302 
\bibitem[McKee \& Tan(2003)]{mt03} McKee, C. F. \& Tan, J. C. 2003, \apj,
       585, 859
\bibitem[McKee \& Zweibel(1995)]{mz95} McKee, C. F. \& Zweibel, E. G. 1995,
       \apj, 440, 686
\bibitem[Ossenkopf \& Mac~Low(2002)]{om02} Ossenkopf, V. \& Mac~Low, M.-M.,
       2002, \aap, 390, 307
\bibitem[Palla \& Stahler(1999)]{ps99} Palla, F. \& Stahler, S. W. 1999,
       \apj, 525, 772 
\bibitem[Palla \& Stahler(2000)]{ps00} Palla, F. \& Stahler, S. W. 2000,
       \apj, 540, 255
\bibitem[Palla \& Stahler(2001)]{ps01} Palla, F. \& Stahler, S. W. 2001,
       \apj, 553, 299
\bibitem[Palla et al.(2005)]{p05} Palla, F., Randich, S., Flaccomio, E., \&
       Pallavicini, R. 2005, \apjl, 626, L49
\bibitem[Pflamm-Altenburg \& Kroupa(2007)]{pk07} Pflamm-Altenburg, J. \&
       Kroupa, P. 2007, \mnras, 375, 855
\bibitem[Press et al.(1988)]{num88} Press, W. H., Teukolsky, S. A., Vetterling,
       W. T., \& Flannery, B. P. 1988, Numerical Recipes, Cambridge U. Press
\bibitem[Pudritz(1990)]{p90} Pudritz, R. E. 1990, \apj, 350, 195
\bibitem[Scalo(1998)]{s98} Scalo, J. 1998, in The Stellar Initial Mass Function,
       ed. G. Gilmore, I. Parry, \& S. Ryan, Cambridge: Cambridge U. Press,
       32
\bibitem[Schmidt(1959)]{s59} Schmidt, M. 1959, \apj, 129, 243
\bibitem[Shirley et al.(2003)]{sh03} Shirley, Y. L., Evans, N. J., Young,
K. E., Knez, C., \& Jaffe, D. T. 2003, \apjs, 149, 375
\bibitem[Stahler \& Palla(2004)]{sp04} Stahler, S. W. \& Palla, F.
       2004, The Formation of Stars, Wiley-VCH 
\bibitem[Stahler et al.(2000)]{sp00} Stahler, S. W., Palla, F., \& Ho, P. T. P.
       2000, in Protostars and Planets IV, ed. V. Mannings, A. P. Boss, \& 
       S. S. Russell, Tucson: U. of Arizona Press, 327
\bibitem[Tan et al.(2006)]{tan06} Tan, J. C., Krumholz, M. R., \&
       McKee, C. F. 2006, \apj, 641, L121
\bibitem[V\'azquez-Semadeni et al.(2003)]{v03} V\'azquez-Semadeni, E., 
Ballesteros-Paredes, J., \& Klessen, R. S. 2003, \apjl, 585, L131
\bibitem[Williams et al.(1995)]{w95} Williams, J. P., Blitz, L., \& Stark, 
       A. A. 1995, \apj, 451, 252
\bibitem[Wolfire et al.(1995)]{wo95} Wolfire, M. G., Hollenbach, D., McKee, 
       C. F., Tielens, A.G.G.M., \& Bakes, E.L.O. 1995, \apj, 443, 152 
\bibitem[Zuckerman \& Evans(1974)]{ze74} Zuckerman, B. \& Evans, N.J. 1974,
      \apjl, 192, L149
\end{thebibliography}
\end{document}